\documentclass[prl,twocolumn,floatfix,showpacs]{revtex4}
\usepackage{graphicx}
\usepackage{mathrsfs}
\usepackage{amsmath}
\usepackage{amssymb}
\usepackage{amsfonts}
\usepackage[normalem]{ulem}
\makeatletter

\begin{document}

\title{Unconventional Spin Density Waves in Dipolar Fermi Gases}
\author{S.~G.~ Bhongale$^{1,4}$, L.~Mathey$^{2}$, Shan-Wen Tsai$^{3}$,
  Charles W. Clark$^{4}$, Erhai Zhao$^{1,4}$} \affiliation{$^{1}$School
  of Physics, Astronomy and Computational Sciences,
  George Mason University, Fairfax, VA 22030\\
  $^{2}$Zentrum f\"ur Optische Quantentechnologien and Institut f\"ur Laserphysik, Universit\"at Hamburg, 22761 Hamburg, Germany\\
  $^{3}$Department of Physics and Astronomy, University of California,
  Riverside, CA 92521 \\
  $^{4}$Joint Quantum Institute, National Institute of Standards and
  Technology \& University of Maryland, Gaithersburg, MD 20899 }
\date{\today}

\begin{abstract} 
The conventional spin density wave (SDW) phase~\cite{overhauser}, as found in
  antiferromagnetic metal for example~\cite{Fawcett}, can be described as a
  condensate of particle-hole pairs with zero angular momentum,
  $\ell=0$, analogous to a condensate of particle-particle pairs in
  conventional superconductors. While many unconventional superconductors
  with Cooper pairs of
  finite $\ell$ have been discovered, their counterparts, density waves with non-zero
  angular momenta, have only been hypothesized in two-dimensional electron
  systems~\cite{nayak}. Using an unbiased functional renormalization group
  analysis, we here show that spin-triplet
  particle-hole condensates with $\ell=1$ emerge generically in 
  dipolar Fermi gases of atoms \cite{benlev} or molecules \cite{jun1,martin}
  on optical lattice.  The order
  parameter of these exotic SDWs is a vector quantity in spin space,
  and, moreover, is defined on lattice bonds rather than on lattice
  sites. We determine the rich quantum phase diagram of dipolar
  fermions at half-filling as a function of the dipolar orientation,
  and discuss how these SDWs arise amidst competition with
  superfluid and charge density wave phases.
\end{abstract}

\maketitle
\begin{figure*}[t]
  \includegraphics[scale=0.19]{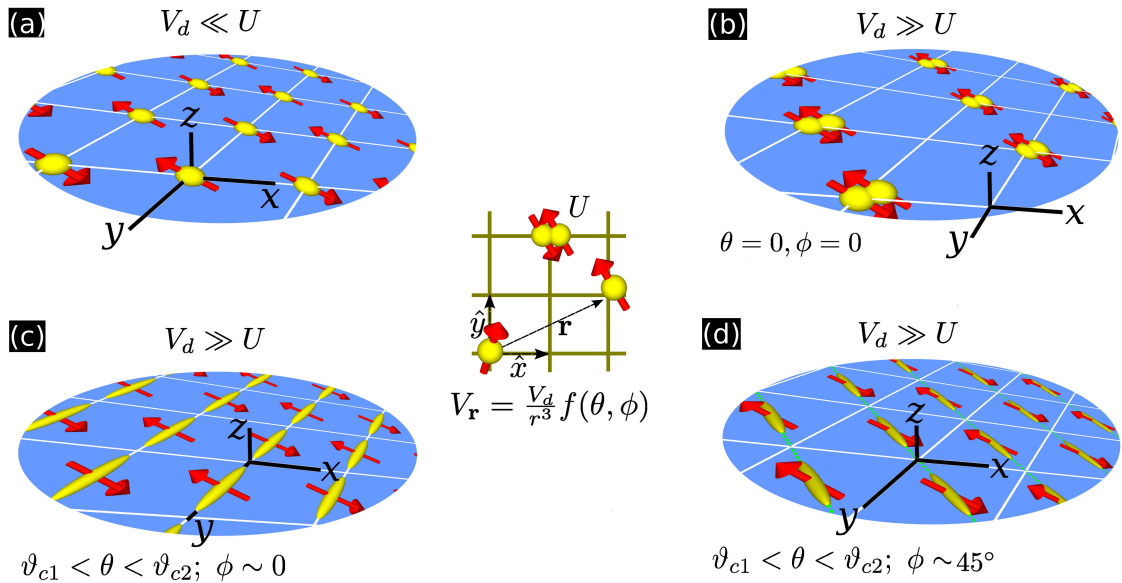}
  \caption{
\textbf{Schematic spin and charge order for pseudo-spin 1/2 dipolar
      fermions in two dimensions}.  
The central image illustrates the
    onsite interaction $U$ between opposite spins (red arrows) and the dipolar
    interaction $V_{\bf{r}}$, here assumed to be spin-independent. The characteristic scale 
    for $V_{\bf{r}}$ is $V_d= d^2/a^3$, with $a$ the lattice spacing. $V_{\bf{r}}$ 
    depends also sensitively on the orientation of the dipoles labelled by angles $(\theta,\phi)$.  
    (a) The conventional antiferromagnetic spin density wave (SDW$_s$). (b) Checkerboard
    charge density wave (CDW$_s$), where ``charge'' is defined as
    the total density. (c) An example of $p$-wave spin density waves
    (SDW$_p$) with modulation of $y$ bond variables.  (d) An example of 
    mixed (extended) $s$- and $d$-wave spin
    density waves (SDW$_{s+d}$).
	Red arrows in (c) and (d) indicate the direction of the spin vector $\pmb{{\cal S}}$ 
    defined on the bonds (yellow ellipsoids).}
\label{af_orders}
\end{figure*}

 
The advent of ultra-cold atomic and molecular
gases has opened new avenues to study many-body physics.  One of the
central subjects of condensed matter physics is quantum magnetism, a
phenomenon that has intrigued scientists for decades.  A
quintessential example is the square-lattice Fermi-Hubbard model at
half-filling, which, even at weak coupling, exhibits SDW, with the
well-known checkerboard pattern depicted in
Fig.~\ref{af_orders}(a). Interestingly, the theory of such
antiferromagnetic order can be cast completely analogous to that of
$s$-wave superconductivity -- with condensation of particle-hole pairs
corresponding to condensation of Cooper pairs in the BCS
theory~\cite{nayak}. This analogy is quite robust and extends to
particle-hole pairing with higher angular momentum, predicting the
existence of a whole array of SDW states, SDW$_{\nu}$, where the label
$\nu=s,p,d,..$ indicates $\ell=0,1,2,..$ respectively
\cite{nayak}. The familiar antiferromagnetic order of
Fig.~\ref{af_orders}(a) constitutes SDW$_s$, whereas SDW$_p$ is the
particle-hole analogue of the spin-triplet $p$-wave superconductors
and superfluid $^3$He \cite{leggett}. In the charge sector, a similar
analogy predicts the existence of generalized $\nu$-wave charge
density wave (CDW$_{\nu}$).

While several candidate systems have been discussed for spin-singlet
charge density waves (CDW$_{\nu\ne s}$) \cite{nayak,bhongale}, up to now the
realization of spin-triplet SDW$_{\nu\ne s}$ has remained elusive.
These exotic states display a modulation of spin vector $\pmb{{\cal S}}$ defined
on the bonds of the lattice, such as the checkerboard pattern 
of red arrows depicted in Figs.~\ref{af_orders}(c)
and \ref{af_orders}(d), as opposed to modulation of on-site variables in
conventional SDW$_s$ and CDW$_s$ shown in
Figs.~\ref{af_orders}(a) and \ref{af_orders}(b). Specifically, 
the
order parameter of SDW$_{p}$, featuring particle-hole
pairs with $p_y$-orbital symmetry, is related to
${\cal S}^\eta =\langle\hat{a}_{i,\alpha}^{\dagger}
\hat{\sigma}^\eta_{\alpha\beta}\hat{a}_{j,\beta}\rangle$
for 
relative coordinate ${\bf r}_{i}-{\bf r}_j=\hat{y}$ (all distances are in units of 
the lattice constant throughout this paper, and repeated indices are summed over).
Here $\hat{\sigma}^{\eta}$ with $\eta \in\{ x, y, z\}$ are the Pauli
matrices, and $\hat{a}_{i,\alpha}^{(\dagger)}$ is the fermionic
annihilation (creation) operators for pseudo-spin $\alpha \in
\{\uparrow, \downarrow\}$ at site $i$. The SDW$_{s+d}$
shown in Fig.~\ref{af_orders}(d) contains an extended $s$-wave and a $d_{xy}$
wave component, and its order parameter is defined similarly with $(i,j)$ corresponding
to diagonally opposite sites, e.g., ${\bf r}_{i}-{\bf
  r}_j=\hat{x}+\hat{y}$.  In contrast, the conventional SDW$_s$ and
CDW$_s$ are described by on-site order parameters
$\langle\hat{a}_{i,\alpha}^{\dagger}
\hat{\sigma}^\eta_{\alpha\beta}\hat{a}_{i,\beta}\rangle$ and $\langle
\hat{n}_i\rangle=\sum_{\alpha}\langle
\hat{n}_{i,\alpha}\rangle=\sum_{\alpha}\langle\hat{a}_{i,\alpha}^{\dagger}\hat{a}_{i,\alpha}\rangle$
respectively.  

The key insight of this paper is that fermions in a 2D lattice with
dominant dipole-dipole interaction have the right ingredients to
stabilize $p$- and $d$-wave SDWs. They emerge between the CDW and the
BCS regime in the phase diagram, as a result of the competition
between the short-ranged inter-atomic and the anisotropic long-ranged
dipolar interaction.

In a new generation of experiments, ultra-cold gases of dipolar
fermions have become accessible in the quantum degenerate limit.
Fermionic atoms of dysprosium 161, with a large magnetic moment of 10 Bohr 
magneton, have been successfully trapped and cooled well below quantum degeneracy
 \cite{benlev}.  The fermionic polar molecule $^{40}$K$^{87}$Rb
has been cooled near quantum degeneracy \cite{jun1} and loaded into
optical lattices. Recently, the formation of ultra-cold fermionic
Feshbach molecules of $^{23}$Na$^{40}$K has been achieved
\cite{martin}. On the theory side, many body physics of single-species
(spinless) dipolar Fermi gases have been explored by many groups.
Numerous quantum phases are predicted: charge density wave
\cite{miyakawa, freericks,parish,gadsbolle}, $p$-wave superfluid
\cite{baranov2, baranov3,taylor,cooper1,hanpu}, liquid crystalline
\cite{fregoso,quin,congjun1}, supersolid \cite{hofstetter}, and
bond-order solid \cite{bhongale}.

Here we consider a two-component (pseudo-spin $1/2$) dipolar Fermi gas
\cite{fradkin,sogo,congjun2} in an optical square lattice at half
filling.  The two pseudo-spin states can be two hyperfine states of Dy
atoms, or two rovibrational states of KRb molecules.  This provides a
tunable platform for quantum simulation of interacting fermions with
long-range interactions \cite{simu,rey}, beyond the Fermi-Hubbard
model. The system is described by the Hamiltonian
\begin{equation}
\hat{H}=-\hspace{-0.2cm}\sum_{\langle i,j\rangle,\sigma} \hspace{-0.1cm}t \hat{a}_{j,\sigma}^{\dagger}
\hat{a}_{i,\sigma} +\frac{U}{2}\sum_{i,\sigma} \hat{n}_{i,\sigma}\hat{n}_{i,-\sigma}
+\sum_{i\ne j}V_{ij}\hat{n}_{i}\hat{n}_{j}.\label{hamiltonian}
\end{equation} 
The lattice is aligned along the $x$- and $y$-directions, with 
nearest neighbor hopping $t$ and on-site
interaction $U$.   $U$
contains contributions from the bare short range interaction, and the
on-site dipolar interaction $V_{ii}^{\perp}$, defined below.  We
assume that all dipoles are aligned in the same direction ${\bf
  d}=d\hat{d}=(d,\theta,\phi)$ by an external magnetic (or electric)
field. In general, the off-site dipole-dipole interaction can be
decomposed into equal- and unequal-spin components, labeled by
$\parallel$ and $\perp$, respectively,
$V_{ij}^{\parallel}\hat{n}_{i\sigma}\hat{n}_{j,\sigma} +
V_{ij}^{\perp}\hat{n}_{i\sigma}\hat{n}_{j,-\sigma}$, and depends on
$\hat{d}$ and ${\bf r}={\bf r}_i-{\bf r}_j$
via  
$V_{{\bf r}}^{\parallel(\perp)}(\hat{d})\equiv
V_{ij}^{\parallel(\perp)}(\hat{d})=\langle
ij|V_{dd}^{\parallel(\perp)}(\hat{d})|ij\rangle=V_d^{\parallel(\perp)}[1-3(\hat{r}\cdot\hat{d})^2]/r^3$.
We will mostly assume
$V_{d}^{\perp}(\hat{d})=V_{d}^{\parallel}(\hat{d}) \equiv V_d(\hat{d})$, as
in Eq. \ref{hamiltonian}, which arises naturally when the two states
are associated with the same hyperfine manifold. 
The $V_{d}^{\perp}(\hat{d})\neq V_{d}^{\parallel}(\hat{d})$ case will be
discussed briefly further down.

\begin{figure*}[t]
  \includegraphics[scale=0.3]{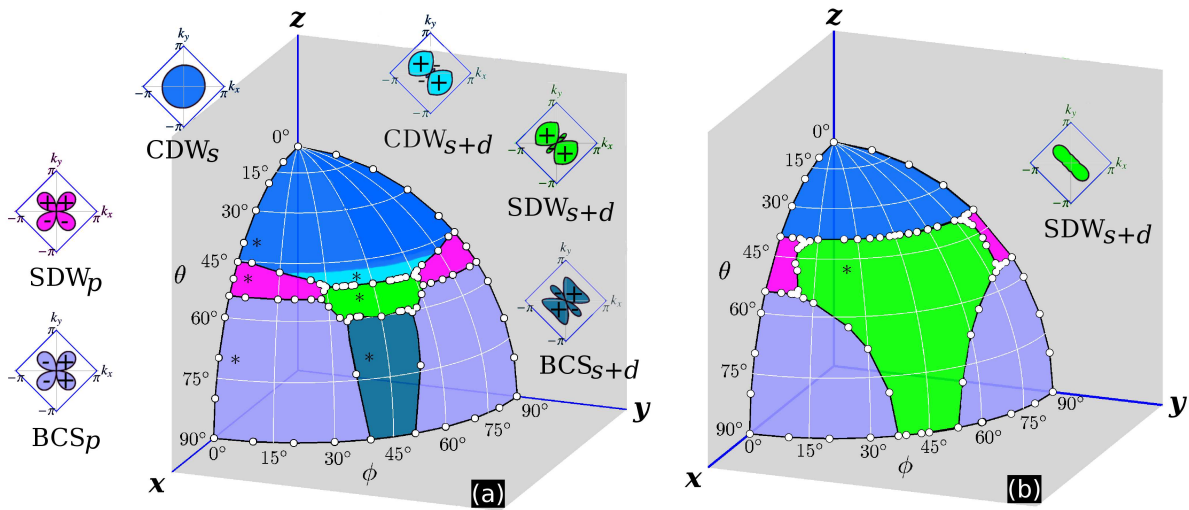}
  \caption{
{\bf Phase diagram of dipolar fermions on a square
      lattice at half filling obtained from FRG.} 
It is shown on the
    surface of a sphere as a function of the dipole orientation angle 
    $\theta$ and $\phi$ for fixed interactions (a) $V_{d}=0.5, U=0.1$; and (b)
    $V_{d}=0.5, U=0.5$ in units of $t$. Unconventional SDWs (SDW$_{p}$ and SDW$_{s+d}$) are 
    sandwiched between the CDW and BCS superfluid phase.   
     The FRG eigenwavefunction corresponding to a representative
    point (marked by
    ``$*$'') in each 
    phase is shown in the $k_x-k_y$ polar plots with
    matching colors. Note that FRG predicts a mixed $s+d$-wave (rather than pure $d$-wave) SDW. 
    As the on-site interaction is increased from $U=0.1$ in (a) to
    $U=0.5$ in (b), the SDW$_{s+d}$ phase expands and squeezes out
    the neighboring phases. The $s$-wave component of
    SDW$_{s+d}$ increases with $U$, while the $d$-wave component diminishes.}
\label{phasedig1}
\end{figure*}

To give a heuristic argument about possible orders of the system, we
consider a simplified version of model (\ref{hamiltonian}) retaining
only the nearest and next-nearest neighbor dipolar interactions, denoted $V_{\hat{x}(\hat{y})}$ and
    $V_{\hat{x}+\hat{y}}$ respectively, see
Fig.~\ref{af_orders}. First, for $\hat{d}=\hat{z}$, dipolar
interactions are purely repulsive.  For $U\gg V_{d}$, the Hamiltonian
reduces to the Fermi-Hubbard model, implying a ground state with
SDW$_s$ order at half-filling, Fig.~\ref{af_orders}(a). For $U\ll
V_{d}$, the dipolar energy is reduced by placing same-spins on
diagonally opposite sites, while opposite spins share the same site
with only a small energy cost $U$. This implies a checkerboard modulation
of the total density $n_i=\langle \hat{n}_i\rangle$, {\it i.e.}
CDW$_s$, shown in Fig.~\ref{af_orders}(b).  

As $\hat{d}$ is tilted away from $\hat{z}$ towards the
$\hat{x}$-direction, there exists a region of tilting direction for
which the nearest neighbor interaction $V_{\hat{x}}$ becomes
attractive while $V_{\hat{y}}$ and $V_{\hat{x}+\hat{y}}$ remain
repulsive.  For instance, for $\phi=0$, this region is bounded by two
critical values of $\theta$:
$\vartheta_{c1}=\cos^{-1}(\sqrt{2/3})\approx 35^{\circ}$ and
$\vartheta_{c2}=\sin^{-1}(\sqrt{2/3})\approx 54^{\circ}$. 
In the simpler case of spinless dipolar fermions, a checkerboard bond order
solid is formed \cite{bhongale} in this region. Then it is plausible 
that for the spin $1/2$ case, unconventional SDWs of non-$s$ wave symmetry may be stabilized
by interaction-induced correlated hopping either along the 
$\hat{x}$, $\hat{y}$, or the diagonal $\hat{x}+\hat{y}$ direction.
The spatial symmetry of these SDWs depends on the value of $\phi$. 
This scenario is illustrated
in  Figs.~\ref{af_orders}(c)
and \ref{af_orders}(d).

Finally, for large dipole tilting angles, {\it e.g.}, $\theta > \vartheta_{c2}$ for $\phi=0$, 
the dominant dipolar interaction is attractive. The leading instability
is towards formation of Cooper pairs. Again, the precise orbital symmetry of the
BCS phase is determined by the value of $\phi$.

We now determine the phase diagram of these intricate competing
orders in the weak coupling limit, $\{U,V_d\}<t$. In particular, we
prove the existence of unconventional SDW phases for intermediate tilting
angles. We use
the functional renormalization group (FRG) technique, which 
takes an unbiased approach (without any {\it a
  priori} guess) to isolate the most dominant instability among
all possible orders \cite{bhongale,shankar,zanchi,mathey,mathey2}. The
FRG used here is an SU(2) symmetric version of that
previously applied to treat spinless dipolar fermions \cite{bhongale}. 
The key ingredients
of the FRG calculation are: (1) Derive and solve the renormalization group flows
for the generalized four-point vertex between unequal spins, ${\cal
  U}_{l}^{\perp}({\bf k}_1,{\bf k}_2,{\bf k}_3)$, where ${\bf
  k}_{1,2}({\bf k}_{3,4})$ are incoming (outgoing) momenta in the
vicinity of the non-interacting Fermi surface, satisfying momentum
conservation ${\bf k_1}+{\bf k_2}={\bf k_3}+{\bf k_4}$, and $l$ is the
renormalization group flow parameter. The flow of equal spin vertex,
${\cal U}_{l}^{\parallel}$, is related to that of ${\cal
  U}_{l}^{\perp}$ via the spin-rotation symmetry of $\hat{H}$. (2) We
project out the interaction channels of interest at each RG step,
\begin{eqnarray}
&{\cal U}_{l}^{\text{CDW}}({\bf k}_1,{\bf
  k}_2)=&(2-\hat{X})\,{\cal U}_{l}^{\perp}({\bf k}_1,{\bf k}_2,{\bf k}_1+{\bf Q}),\nonumber\\ 
&{\cal U}_{l}^{\text{SDW}}({\bf k}_1,{\bf
  k}_2)=&-\hat{X}{\cal U}_{l}^{\perp}({\bf k}_1,{\bf k}_2,{\bf k}_1+{\bf Q}),\nonumber\\
 &{\cal U}_{l}^{\text{BCS}}({\bf k}_1,{\bf
  k}_2)=&{\cal U}_{l}^{\perp}({\bf k}_1,-{\bf k}_1,{\bf k}_2,-{\bf k}_2),\nonumber
\end{eqnarray}
where the exchange operator $\hat{X}$ interchanges the incoming
momenta. (3) Finally we identify the most dominant instability of the
Fermi surface from the most divergent eigenvalue of the interaction
matrix. The corresponding eigenvector provides information about the
orbital symmetry of the incipient order parameter.
The phase diagram is shown in
Fig.~\ref{phasedig1}.

 \begin{figure}[ht]
  \includegraphics[scale=.55]{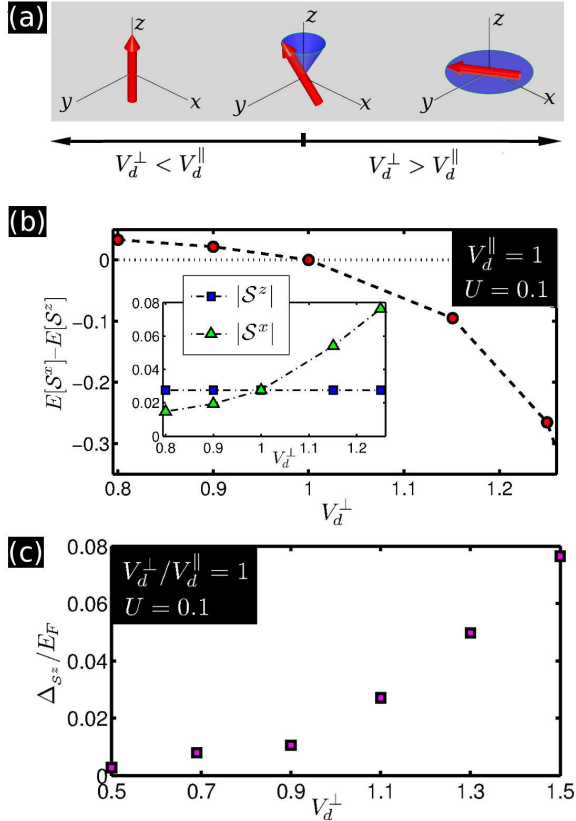}
  \caption{
{\bf Mean field results for the SDW$_p$ phase.} 
(a) Preferred direction of the
    spin polarization vector $\pmb{\cal S}$ as a function of the ratio
    $V_d^{\parallel}/V_d^{\perp}$. It is along the z-direction for
    $V_{d}^{\perp}<V_d^{\parallel}$ and lies in the $x$-$y$ plane for
    $V_{d}^{\perp}>V_d^{\parallel}$. (b) The energy difference between the
    mean field states with ${\cal S}^z$ and ${\cal S}^x$ order. The
    inset shows the magnitude of the corresponding order
    parameter. The parameters are $U=0.1$, $V_{d}^{\parallel}=1.0$,
    $\theta=47^\circ$, $\phi=0$. All energies are in units of $t$.
    (c) The mean field energy gap of the SDW$_p$ phase, in units of the Fermi
    energy, as a function of the inter-spin long-range 
    dipolar interaction $V_d^{\perp}$ for
    $\theta=47^{\circ}$ and $\phi=0$. For the SU(2)
    symmetric case plotted in (c), the energy gaps for different vector
    polarizations are degenerate.}
\label{energy_compare}
\end{figure}

The phase diagram displays three types of phases: CDW, SDW, and BCS
superfluid. We first focus on the case $U<V_{d}$ in the
vicinity of $\phi=0$ as shown in Fig.~\ref{phasedig1}(a). Consistent
with our heuristic argument above, FRG confirms a checkerboard CDW
(CDW$_s$) for small $\theta$, and a spin-triplet, $p$-wave BCS
(BCS$_p$) superfluid at large $\theta$.  For the 
intermediate regime, roughly between $\vartheta_{c1}$ and
$\vartheta_{c2}$, the flow for the SDW channel diverges rapidly,
dominating over the CDW and BCS instabilities on either side. The
SDW phase shows $p$-wave orbital symmetry, i.e. the eigenvector of the
SDW$_p$ phase (shown in Fig.~\ref{phasedig1}) is essentially of the
form $\sin k_y$. This admits an interpretation of SDW$_p$ as a
particle-hole analog of triplet superconductivity/superfluidity within
Nayak's classification \cite{nayak} for generalized SDW$_{\nu}$.  The
SDW$_p$ phase found here corresponds to the class with
$\langle \hat{a}^{\dagger}_{\alpha}(\mathbf{k}+\mathbf{Q})
\hat{a}_{\beta}(\mathbf{k})\rangle=\mathbf{S}(\mathbf{k})\cdot
\boldsymbol{\sigma}_{\alpha\beta}$, by identifying
$\mathbf{S}(\mathbf{k})\propto \hat{s} \sin k_y$ where
$\mathbf{Q}=(\pm\pi,\pm\pi)$.  The position space representation implies
the checkerboard pattern of hopping amplitudes,
$\langle \hat{a}^{\dagger}_{i,\alpha}
\hat{a}_{j,\beta}\rangle$; ${\bf r}_j-{\bf r}_i=\hat{y}$, depicted in the schematic of
Fig.~\ref{af_orders}(c).

Additional unconventional orders with $\ell \neq 0$ occur in the vicinity
of $\phi=45^{\circ}$, where the nearest-neighbor interaction along the
lattice vectors $\hat{x}$ and $\hat{y}$ is nearly equal. 
FRG predicts three more phases, CDW$_{s+d}$, SDW$_{s+d}$, and BCS$_{s+d}$,
all of which contain a $d_{xy}$-wave as well as $s$-wave components.
The contributions of the isotropic $s$-wave, extended $s$-wave, and $d$-wave
components are inferred by fitting the FRG wavefunctions using 
function $c_0+c_1 \cos k_x\cos k_y+c_2 \sin k_x\sin k_y$, with
$\{c_0,c_1,c_2\}$ as fitting parameters. As a general trend, for increasing $\theta$,
the magnitude of isotropic $s$-wave $c_0$ reduces, while the magnitudes of 
$c_1$ and $c_2$ are comparable and increase.
The CDW$_{s+d}$ phase can be viewed as the natural
continuation of the CDW$_s$ as $c_1$ and $c_2$ become appreciable.
The two representative
points shown in Fig.~\ref{phasedig1}(a) for the SDW$_{s+d}$ and BCS$_{s+d}$
are fit by $0.05 - 0.16 \cos k_x \cos k_y - 0.18 \sin k_x \sin k_y$ and 
$0.01+ 0.23 \cos k_x \cos k_y - 0.19 \sin k_x \sin k_y$, respectively.
Since $c_0$ is small, the real space modulation pattern for such 
SDW$_{s+d}$ takes the form of Fig~\ref{af_orders}(d): atoms delocalize across a
plaquette, in the diagonal direction perpendicular to the dipole tilting direction.
In contrast to the triplet BCS$_p$
phase at small $\phi$, the BCS$_{s+d}$ phase is a superfluid of singlet Cooper pairs 
with mixed orbital symmetry, $\ell=0,2$.

Next we illustrate how the phase diagram changes as the model approaches the
repulsive Fermi-Hubbard model $(U> 0, V_d=0)$. We
calculate the FRG flows for increased on-site interaction, $U=0.5$,
while keeping $V_d$ fixed at $0.5$. The phase diagram is shown in
Fig.~\ref{phasedig1}(b). Since the on-site interaction $U$ favors
antiferromagnetism, the SDW$_p$ phase shrinks, while the SDW$_{s+d}$
phase extends to cover a broader region, including that previously
occupied by BCS$_{s+d}$. Note that the $d$-wave component of SDW, even
though diminished, is always present since the dipole interaction is
kept finite. When $U$ is further increased such that $U\gg
V_d$, only the isotropic component ($c_0$) will survive, indicating 
the SDW$_{s+d}$ crosses over to SDW$_s$,
the conventional
antiferromagnetic ordering of spins in Fig.~\ref{af_orders}(a).

To corroborate the FRG prediction of the unconventional spin density
waves, we use self-consistent mean field theory.
For a square lattice of
finite size $N\times N$, we impose periodic boundary conditions and
retain the dipole interactions up to a distance of 12 lattice
constants. We define the various normal and anomalous averages,
$\rho_{i,\sigma, j,\sigma'}=\langle
\hat{a}_{j,\sigma'}^{\dagger}\hat{a}_{i,\sigma}\rangle$ and
$m_{i,\sigma, j,\sigma'}=\langle
\hat{a}_{i,\sigma}\hat{a}_{j,\sigma'}\rangle$. The corresponding mean
field Hamiltonian is solved self-consistently by starting from an
initial guess of the generalized density matrix, and iterating until
desired convergence is reached. At each step the chemical potential is
tuned to maintain half filling. The results are checked to be
size-independent by varying $N>24$. In Fig.~\ref{energy_compare}, $N$ 
is set to 28. Although mean field results are only suggestive, they provide an
independent confirmation of the FRG results and unveil the real space
patterns of $\pmb{\cal S}$ directly in the SDW$_\nu$ phases. 
They can also be used to investigate the direction of $\pmb{\cal S}$
for the generalized model with $V_{d}^{\perp}(\hat{d})\neq V_{d}^{\parallel}(\hat{d})$.
We search
for unconventional SDW phases with homogeneous spin density,
$n_{i,\sigma}=1/2$. In and around the SDW$_p$ region predicted by FRG,
we indeed find solutions with order parameter ${\cal S}^\eta=\langle
\hat{a}_{i,\alpha}^{\dagger}
\hat{\sigma}^\eta_{\alpha\beta}\hat{a}_{j,\beta}\rangle$, ${\bf
  r}_{j}-{\bf r}_i=\hat{y}$. Further, the mean field energy for
 ${\cal S}^x$ order is identical to that for ${\cal S}^y$ and ${\cal S}^z$, 
 due to the SU(2) symmetry of $\hat{H}$ imposed by
$V_d^{\perp}(\hat{d})=V_d^{\parallel}(\hat{d})$.
This degeneracy is
lifted for $V_d^{\perp}(\hat{d})\ne V_d^{\parallel}(\hat{d})$.  In
Fig.~\ref{energy_compare}, we compare the mean-field energies of the
SDW$_p$ solution with order parameter ${\cal S}^z$ and ${\cal
  S}^x$. The $z(x)$- polarized order ${\cal S}^z$ (${\cal S}^x$) is
energetically favored for $V_{d}^{\parallel}> V_{d}^{\perp}$
($V_{d}^{\parallel}< V_{d}^{\perp}$). However, the degeneracy between
${\cal S}^x$ and ${\cal S}^y$ remains. The mean field results support our
interpretation of the SDW$_p$ order as schematically shown in
Fig.~\ref{af_orders}(c). A similar analysis can be performed for the
SDW$_{s+d}$ phase.

In conclusion, we have established the emergence of unconventional
spin density wave orders, SDW$_p$ and SDW$_{s+d}$, along with other exotic
phases with non-zero angular momentum, within ultra-cold
spin-$1/2$ dipolar fermions on the square lattice. These
phases occupy a sizable region of the phase diagram mapped out via the
functional renormalization group approach. Furthermore,
the self-consistent mean field estimation for the energy gaps of SDW$_p$, shown in
Fig.~\ref{energy_compare}(c), indicates a critical
temperature $T_c \sim 0.08 T_F$ for $V_d/t=1.5$. Considering the
currently reported temperature of degenerate dysprosium $T\approx 0.2
T_F$ \cite{benlev}, this suggests experimental accessibility of these
emergent phases in the near future.  Our study thus articulates the
dipolar Fermi system as an intriguing and novel test bed for exotic
many-body effects. They provide a fresh perspective on systems with
competing orders.

We thank Benjamin Lev for invaluable 
discussions. SB and EZ are supported by AFOSR 
(FA9550-12-1-0079) and ONR (N00014-09-1-1025A). LM acknowledges support from the
Landesexzellenzinitiative Hamburg, which is financed by the Science
and Research Foundation Hamburg and supported by the Joachim Herz
Stiftung.  SWT acknowledges support from NSF under grant DMR-0847801
and from the UC-Lab FRP under award number 09-LR-05-118602.

\end{document}